\documentclass[preprintnumbers,amsmath,amssymbm,prd]{revtex4}
\usepackage{epsfig}
\usepackage{graphicx}
\usepackage{amssymb}

\begin{document}

\title{On the maximum force conjecture in curved spacetimes of stable self-gravitating matter configurations}
\author{Shahar Hod}
\affiliation{The Ruppin Academic Center, Emeq Hefer 40250, Israel}
\affiliation{ } \affiliation{The Hadassah Institute, Jerusalem
91010, Israel}
\date{\today}

\begin{abstract}

\ \ \ Gibbons and Schiller have raised the physically interesting conjecture that forces in general relativity 
are bounded from above by the mathematically compact relation ${\cal F}\leq c^4/4G$. 
In the present compact paper we explicitly prove, using the non-linearly coupled 
Einstein-matter field equations, that the force function ${\cal F}\equiv 4\pi r^2 p(r)$ in {\it stable} 
self-gravitating horizonless matter configurations is characterized by the 
upper bound ${\cal F}\leq c^4/G$ [here $p(r)$ is the 
radial pressure inside the self-gravitating matter configuration]. 
\end{abstract}
\bigskip
\maketitle

\section{Introduction}

The maximum force conjecture in general relativity was raised 
by Gibbons \cite{Gib} and Schiller \cite{Sch1,Sch2} more than two decades ago. 
This intriguing conjecture asserts that, within the framework of general relativity, forces 
are bounded from above by a classical (no-$\hbar$) relation of the form
\begin{equation}\label{Eq1}
{\cal F}\leq \eta\cdot{{c^4}\over{G}}\  ,
\end{equation}
where $\eta$ is a dimensionless factor. 
In particular, the strong version of the maximum force conjecture asserts that $\eta=1/4$ in 
the inequality (\ref{Eq1}) \cite{Gib,Sch1}, whereas $\eta=O(1)$ in 
the weak version of the conjecture \cite{Ong}. Interestingly, it was claimed \cite{Sch3} 
that the maximum force 
relation (\ref{Eq1}) can be used to derive the Einstein-matter field equations of general relativity. 

In the present compact paper we raise the following physically important 
question: Can a maximum force relation of the form (\ref{Eq1}) be derived in a mathematically rigorous way for 
spatially regular self-gravitating matter configurations in curved spacetimes?

Interestingly, below we shall explicitly prove that the non-linearly coupled Einstein-matter field equations, 
supplemented by the physically motivated requirement of dynamical {\it stability}, yield an upper bound on the 
radially-dependent dimensionless force function \cite{NoteUnits}
\begin{equation}\label{Eq2}
{\cal F}\equiv 4\pi r^2\cdot p(r)\
\end{equation}
in self-gravitating matter configurations. 

In particular, we shall explicitly show below that the existence of an upper bound on the magnitude of the 
force function (\ref{Eq2}) in curved spacetimes of self-gravitating field configurations is closely related 
to the mathematically elegant theorem presented in \cite{CBH} (see also \cite{Hodrw,Hoddd}), according to
which the innermost light ring (the innermost closed null circular geodesic) of an 
horizonless curved spacetime, if it exists, is stable \cite{Notesth}. 

As nicely discussed in \cite{Keir,HerNew} (see also \cite{Hodt1}), the presence of stable light rings in 
spatially regular spacetimes of self-gravitating matter configurations implies that the corresponding curved 
spacetimes develop non-linear instabilities in the presence of massless fields \cite{Notekk}. 
One therefore concludes that a necessary condition for self-gravitating matter 
configurations to be dynamically {\it stable} is that the corresponding curved spacetimes 
do not possess null circular geodesics (closed light rings). 

Motivated by the physically intriguing observation made in
\cite{CBH,Keir,HerNew} regarding the (in)stability properties of spatially regular 
self-gravitating matter configurations in classical general relativity, 
in the present paper we shall use the non-linearly coupled Einstein-matter field equations in order 
to derive an explicit upper bound on the dimensionless force function (\ref{Eq2}) in 
{\it stable} curved spacetimes \cite{JoVi,NoteJV1,NoteJV2}. 

\section{Description of the system}

We consider spatially regular self-gravitating matter configurations whose curved 
spacetimes are described by the spherically symmetric line element
\cite{Chan,ShTe,Hodt1,Hodt2,Notesc}
\begin{equation}\label{Eq3}
ds^2=-e^{-2\delta}\mu dt^2 +\mu^{-1}dr^2+r^2(d\theta^2 +\sin^2\theta d\phi^2)\  ,
\end{equation}
where $\mu=\mu(r)$ and $\delta=\delta(r)$. 

The non-linearly coupled Einstein-matter field equations $G^{\mu}_{\nu}=8\pi T^{\mu}_{\nu}$ of the 
spherically symmetric spacetime (\ref{Eq3}) can be expressed in the form 
\cite{Chan,BekMay,Hodt1,Noteprm}
\begin{equation}\label{Eq4}
\mu'=-8\pi r\rho+{{1-\mu}\over{r}}\
\end{equation}
and
\begin{equation}\label{Eq5}
\delta'=-{{4\pi r(\rho +p)}\over{\mu}}\  ,
\end{equation}
where $\rho\equiv -T^{t}_{t}$ and $p\equiv T^{r}_{r}$ are respectively the energy density and the 
radial pressure that characterize the matter fields \cite{Bond1}.

The metric functions of spatially regular asymptotically flat spacetimes are characterized 
by the boundary relations \cite{BekMay,Hodt1}: 
\begin{equation}\label{Eq6}
\mu(r\to 0)\to1\ \ \ \ ; \ \ \ \ \delta(r\to0)<\infty\
\end{equation}
and
\begin{equation}\label{Eq7}
\mu(r\to\infty)\to 1\ \ \ \ ; \ \ \ \ \delta(r\to\infty)\to 0\  .
\end{equation} 
In addition, horizonless matter configurations, which are the focus of the present study, 
are characterized by the inequality 
\begin{equation}\label{Eq8}
\mu(r)>0\ \ \ \ \text{for}\ \ \ \ r\in[0,\infty]\  .
\end{equation}

Taking cognizance of the Einstein equation (\ref{Eq4}) one finds that the dimensionless 
metric function $\mu(r)$ can be expressed in the compact mathematical form \cite{BekMay,Hodt1}
\begin{equation}\label{Eq9}
\mu(r)=1-{{2m(r)}\over{r}}\  ,
\end{equation}
where 
\begin{equation}\label{Eq10}
m(r)=\int_{0}^{r} 4\pi x^2\rho(x)dx\
\end{equation}
is the radially-dependent gravitational mass contained within a sphere of 
radius $r$ \cite{BekMay,Hodt1}. 

We shall assume that the spatially regular self-gravitating matter configurations are characterized by 
the dominant energy condition,  
\begin{equation}\label{Eq11}
0\leq |p|\leq\rho\  ,
\end{equation}
which implies the relation [see Eqs. (\ref{Eq9}) and (\ref{Eq10})]
\begin{equation}\label{Eq12}
\mu(r)\leq1\ \ \ \ \text{for}\ \ \ \ r\in[0,\infty]\
\end{equation}
for the dimensionless metric function. 

For later purposes we note that Eq. (\ref{Eq10}) implies that 
asymptotically flat spacetimes (with finite ADM masses) are characterized by the asymptotic 
radial behavior 
\begin{equation}\label{Eq13}
r^3\cdot \rho(r)\to 0\ \ \ \ \text{for}\ \ \ \ r\to\infty\  ,
\end{equation}
which also implies the asymptotic relation [see Eq. (\ref{Eq11})]
\begin{equation}\label{Eq14}
r^3\cdot p(r)\to 0\ \ \ \ \text{for}\ \ \ \ r\to\infty\
\end{equation}
for the pressure function in the curved spacetime (\ref{Eq3}). 

\section{Maximum force relation in curved spacetimes of 
stable spatially regular self-gravitating matter configurations}

In the present section we shall explicitly prove that one can use the instability properties of horizonless spacetimes that 
possess null circular geodesics (closed light rings) \cite{CBH,Hodrw,Hoddd,Keir,HerNew,Hodt1} 
in order to derive an upper bound on the dimensionless force function (\ref{Eq2}) that characterizes {\it stable} 
spatially regular self-gravitating matter configurations. 

The radial functional relation that determines the radii of null circular geodesics in
the curved spacetime (\ref{Eq3}) was derived in
\cite{Chan,NoteHodt1}. 
Before we present our proof for the existence of an explicit upper bound on the dimensionless 
force function (\ref{Eq2}) in curved spacetimes of stable self-gravitating matter configurations, 
we shall first provide here a brief sketch of the analytical derivation 
of the characteristic equation [see Eq. (\ref{Eq17}) below] that 
determines the locations of closed light rings in spherically symmetric curved spacetimes. 

As explicitly shown in \cite{Chan,NoteHodt1}, null circular geodesics are characterized by
the functional relations \cite{Notethr,Notedot}
\begin{equation}\label{Eq15}
V_r=E^2\ \ \ \ \ \text{and}\ \ \ \ \ V'_r=0\  ,
\end{equation}
where the effective radial potential $V_r$ is determined by the expression \cite{Chan,NoteHodt1}
\begin{equation}\label{Eq16}
E^2-V_r\equiv \dot
r^2=\mu\Big({{E^2}\over{e^{-2\delta}\mu}}-{{L^2}\over{r^2}}\Big)\  .
\end{equation}
Here the physical parameters $\{E,L\}$ are respectively the energy 
and the angular momentum that characterize the geodesic motion in the curved spacetime. 
These conserved physical quantities 
reflect the fact that the spherically symmetric static spacetime (\ref{Eq3}) has no dependence on the 
time and angular coordinates $\{t,\phi\}$ \cite{Chan,NoteHodt1}.

Substituting (\ref{Eq16}) into the functional relations (\ref{Eq15}) and taking cognizance of the 
Einstein equations (\ref{Eq4}) and (\ref{Eq5}), one
finds that the radial relation
\begin{equation}\label{Eq17}
{\cal R}(r=r_{\gamma})=0\
\end{equation}
with
\begin{equation}\label{Eq18}
{\cal R}(r)\equiv 3\mu(r)-1-8\pi r^2 p(r)\
\end{equation}
determines the radii of null circular geodesics (closed light rings) in the spherically symmetric 
curved spacetime (\ref{Eq3}).

In addition, from Eqs. (\ref{Eq6}), (\ref{Eq7}), (\ref{Eq14}), and (\ref{Eq18}) 
one learns that the dimensionless radial function ${\cal R}(r)$ is characterized by the boundary relations
\begin{equation}\label{Eq19}
{\cal R}(r=0)=2\
\end{equation}
and
\begin{equation}\label{Eq20}
{\cal R}(r\to\infty)\to 2\  .
\end{equation}
Taking cognizance of Eqs. (\ref{Eq17}), (\ref{Eq19}), and (\ref{Eq20}), 
one deduces that horizonless curved spacetimes of spatially regular self-gravitating 
matter configurations are generally characterized by an {\it even} (possibly zero) number of closed 
light rings \cite{CBH,Hoddd,NoteHoddd}.

The (in)stability of null circular geodesics in the curved spacetime (\ref{Eq3}) is 
determined by the second spatial derivative of the radially-dependent 
curvature potential (\ref{Eq16}) \cite{Chan,NoteHodt1}. 
In particular, stable light rings which, as discussed in \cite{Keir,HerNew,Hodt1}, are
associated with non-linear dynamical instabilities of massless fields in the corresponding curved
spacetimes, are characterized by locally convex curvature potentials
with the radial property $V''_r(r=r_{\gamma})>0$ \cite{Chan,NoteHodt1}.

From Eqs. (\ref{Eq4}), (\ref{Eq5}), (\ref{Eq16}), (\ref{Eq17}), and 
the conservation equation $T^{\mu}_{r;\mu}=0$ \cite{BekMay}, one obtains the radial 
functional relation \cite{Hoddd,Hodrw}
\begin{equation}\label{Eq21}
V''_r(r=r_{\gamma})=-{{E^2e^{2\delta}}\over{\mu r_{\gamma}}}\times
{\cal R}'(r=r_{\gamma})\
\end{equation}
for the second radial derivative of the effective curvature potential (\ref{Eq16}). 
Interestingly, taking cognizance of Eqs. (\ref{Eq17}) and (\ref{Eq19}) one finds that, 
in general \cite{NoteHoddd}, the innermost null circular geodesic 
of a spatially regular self-gravitating matter configuration is characterized by the properties
\begin{equation}\label{Eq22}
{\cal R}(r=r^{\text{innermost}}_{\gamma})=0\ \ \ \ \ \text{and}\ \ \ \ \ 
{\cal R}'(r=r^{\text{innermost}}_{\gamma})<0\  .
\end{equation}
From Eqs. (\ref{Eq21}) and (\ref{Eq22}) one deduces that the innermost null circular geodesic 
of a spatially regular self-gravitating matter configuration is {\it stable}. In particular, it is 
characterized by the property \cite{Hodrw,CBH,Hoddd} 
\begin{equation}\label{Eq23}
V''_r(r=r^{\text{innermost}}_{\gamma})>0\  .
\end{equation}

As discussed above, the presence of the stable light ring (\ref{Eq22}) in the 
curved spacetime (\ref{Eq3}) implies that the spatially regular self-gravitating 
matter configuration is non-linearly {\it unstable} to massless perturbation fields
\cite{Keir,HerNew,Notekk}. One therefore concludes that physically realistic (that is, {\it stable}) 
self-gravitating matter configurations have no null circular geodesics (no closed light rings) 
with the functional properties (\ref{Eq22}). 
In particular, taking cognizance of Eqs. (\ref{Eq17}), (\ref{Eq19}), and (\ref{Eq20}), 
one deduces that spatially regular {\it stable} matter configurations 
are characterized by the radial property 
\begin{equation}\label{Eq24}
R(r)>0\ \ \ \ \ \text{for}\ \ \ \ \ r\in[0,\infty]\  ,
\end{equation}
which yields the upper bound [see Eq. (\ref{Eq18})]
\begin{equation}\label{Eq25}
{\cal F}(r)<{1\over2}\big[3\mu(r)-1\big]\ \ \ \ \ \text{for}\ \ \ \ \ r\in[0,\infty]\
\end{equation}
on the dimensionless force function (\ref{Eq2}). 

\section{Summary}

In the physically important papers \cite{Gib,Sch1} Gibbons and Schiller have conjectured that the 
compact mathematical relation ${\cal F} \leq c^4/4G$ provides an upper bound on the magnitude of 
forces in general relativity. 
Motivated by this intriguing conjecture, in the present paper we have raised the physically interesting 
question: Can a maximum force relation of the form ${\cal F}\leq \eta\cdot c^4/G$ with $\eta=O(1)$ 
be derived for spatially regular self-gravitating {\it stable} matter configurations in curved spacetimes?

In order to derive the desired upper bound on the dimensionless force function ${\cal F}\equiv 4\pi r^2 p(r)$, 
we have pointed out that the above stated question is closely related to the recent
theorem presented in \cite{CBH} (see also \cite{Hodrw,Hoddd}) which, when combined with the physically 
interesting results presented in \cite{Keir,HerNew}, asserts that horizonless curved spacetimes of 
spatially regular self-gravitating matter configurations that possess closed light rings (null circular geodesics) 
are dynamically unstable to non-linear massless fields.

Using this characteristic instability property of self-gravitating matter configurations that possess null 
circular geodesics, we have explicitly proved that the answer to the
above stated intriguing question is `Yes!'. 
In particular, using the non-linearly coupled 
Einstein-matter field equations, we have proved that the force function ${\cal F}\equiv 4\pi r^2 p(r)$ in curved 
spacetimes of {\it stable} self-gravitating matter configurations is bounded from above by the remarkably 
compact relation [see Eqs. (\ref{Eq12}) and (\ref{Eq25})]
\begin{equation}\label{Eq26}
{\cal F}<{{c^4}\over{G}}\  .
\end{equation}

\bigskip
\noindent {\bf ACKNOWLEDGMENTS}

This research is supported by the Carmel Science Foundation. I would
like to thank Yael Oren, Arbel M. Ongo, Ayelet B. Lata, and Alona B.
Tea for stimulating discussions.


\begin{thebibliography}{99}
\bibitem{NoteUnits} We shall henceforth use gravitational units with $G=c=1$, 
in which case the radially-dependent force function (\ref{Eq2}) is dimensionless.
\bibitem{Notesth} Note that closed light rings in curved spacetimes are considered to be
stable (attractive) if massless fields tend to pile up on them.
\bibitem{Notekk} As discussed in \cite{Keir,HerNew}, 
this physically interesting assertion is based on the fact that massless fields in curved spacetimes tend to pile up
and grow non-linearly around stable null circular geodesics.
\bibitem{NoteJV2} It is worth noting that it has been shown in \cite{JoVi} that the equatorial force function 
${\cal F}_{\text{eq}}\equiv 2\pi\int^{R_{\text{s}}}_{0}\sqrt{g_{rr}}p_{\text{t}}(r)rdr$ of perfect fluid configurations 
may diverge (here $R_{\text{s}}$ and $p_{\text{t}}$ are respectively the radius of the perfect fluid 
matter configuration and its tangential pressure). 
As nicely emphasized in \cite{JoVi}, it is important to specify precisely which forces are considered. 
In the present analytical study we shall focus our attention on 
the functional behavior of the radial pressure function (\ref{Eq2}). 
In particular, we will prove that, for stable matter configurations, it is bounded from above.
\bibitem{Notesc} Here $(t,r,\theta,\phi)$ are the familiar Schwarzschild coordinates of the curved spacetime.
\bibitem{Noteprm} The prime symbol $'$ is used here to denote a spatial derivative of the 
metric function with respect to the radial coordinate $r$ of the curved spacetime.
\bibitem{Notethr} Note that the radial properties (\ref{Eq15}), which characterize the closed light rings 
of the curved spacetime (\ref{Eq3}), correspond to $\dot r^2=(\dot r^2)'=0$ \cite{Chan}.
\bibitem{Notedot} The dot symbol $^{.}$ is used here to denote a derivative with respect to an affine parameter.
\bibitem{NoteHoddd} See \cite{Hoddd} for the interesting case of
horizonless curved spacetimes with degenerate closed null 
circular geodesics which are characterized by the functional relations ${\cal R}={\cal R}'=0$. 
As discussed in \cite{Hoddd}, these special spacetimes may have an odd
number of closed light rings.
\end{thebibliography}
\end{document}